\documentclass[a4paper]{jpconf}
\usepackage{graphicx}

\def\q{\mbox{\boldmath $q$}}
\begin{document}
\title{Models for quasielastic electron and neutrino-nucleus scattering}

\author{Carlotta Giusti and Andrea Meucci}

\address{Dipartimento di Fisica Nucleare e Teorica, 
Universit\`{a} degli Studi di Pavia and \\
INFN, Sezione di Pavia, via Bassi 6 I-27100 Pavia, Italy}

\ead{Carlotta.Giusti@pv.infn.it}

\begin{abstract}
Models developed for the exclusive and inclusive quasielastic (QE)  
electron-nucleus scattering have been extended to  QE
neutrino-nucleus scattering. Different descriptions of final-state interactions 
(FSI) are compared. For the inclusive electron scattering the relativistic 
Green's function model (RGF) is compared with a model based on the use of 
relativistic purely real mean field (RMF) potentials in the final state.
Both approaches lead to a redistribution of the strength but conserving the 
total flux. Results for electron and neutrino scattering are presented and 
discussed in different conditions and kinematics. The results of the RGF and RMF 
models are compared with the double-differential charged-current QE neutrino 
cross sections recently measured by the MiniBooNE
collaboration using a carbon target.
\end{abstract}

\section{Introduction}
Several decades of experimental and theoretical work on electron scattering have
provided a wealth of information on nuclear structure and 
dynamics~\cite{rep,book}. 
Additional information on nuclear properties is in principle available from 
neutrino-nucleus scattering.  
The main aim of neutrino experiments, however, is not to investigate nuclear
properties, but to determine neutrino properties. In neutrino oscillation 
experiments neutrino-nucleus scattering is used to detect neutrinos and a 
proper analysis of data requires that the nuclear response to neutrino 
interactions is well under control and that the unavoidable theoretical 
uncertainties on nuclear effects are reduced as much as possible.

Different models developed and successfully tested in comparison
with electron-scattering data have been extended to describe nuclear effects in 
neutrino-nucleus scattering. 
Although the two situations are different, electron scattering is the best 
available guide to determine the predictive power of a nuclear model. 
Nonrelativistic and relativistic models have been developed to describe nuclear
effects with different approximations. In general, they can be considered as
different and alternative models. Relativity is however important at all 
energies, in particular at high energies, and in the energy regime of many 
neutrino experiments a fully relativistic approach is required, where not only
relativistic kinematics is considered, but also nuclear dynamics and current 
operators should be described within a relativistic framework. 

Models for the exclusive and inclusive electron and neutrino scattering in the 
QE region are presented in this contribution. In the QE region the nuclear 
response is dominated by the mechanism of one-nucleon knockout, where the probe 
interacts with a quasifree nucleon that is emitted from the nucleus with a 
direct one-step mechanism and the remaining nucleons are spectators.
In electron-scattering experiments the outgoing nucleon can be detected in
coincidence with the scattered electron. In the exclusive $(e,e'p)$ reaction the
residual nucleus is left in a specific discrete eigenstate and the final state 
is completely specified. In the inclusive $(e,e')$ scattering only the scattered
electron is detected, the final nuclear state is not determined, and the cross 
section includes all the available final nuclear states. 

For an incident neutrino (antineutrino) neutral-current (NC) and charged-current
(CC) scattering can be considered  
\begin{eqnarray}
\nu (\bar\nu) + A & \rightarrow & {\nu'} (\bar\nu') + N + 
( A - 1)  \hspace{3cm} \mathrm{NC} \nonumber \\
\nu (\bar\nu) + A & \rightarrow & l^{-} (l^{+}) +
p(n) + (A-1)  \hspace{2.47cm} \ \mathrm{CC} 
 \nonumber\end{eqnarray} 
In NC scattering only the emitted nucleon can be detected and the cross 
section is integrated over the energy and angle of the final lepton. Also 
the state of the residual $(A-1)$-nucleus is not determined and the cross 
section is summed over all the available final states. The 
same situation occurs for the CC reaction if only the outgoing nucleon is 
detected. The cross sections are therefore inclusive in the leptonic sector and 
semi-inclusive in the hadronic sector. The exclusive CC process where the 
charged final lepton is detected in coincidence with the emitted nucleon can be considered as well. 
The inclusive CC scattering where only the charged lepton is detected 
can be treated with the same models used for the inclusive $(e,e')$ reaction. 

For all these processes the cross section is obtained  in the one-boson 
exchange approximation from the contraction between the lepton tensor, that
under the assumption of the plane-wave approximation for the initial and final
lepton wave functions depends  only on the lepton kinematics,  and the hadron 
tensor $W^{\mu\nu}$, that contains the nuclear response and whose components 
are given by bilinear products of the matrix elements of the nuclear current  
$J^{\mu}$ between the initial and final nuclear states, {\it i.e.},
\begin{equation}
W^{\mu\nu} = \sum_f \, \langle \Psi_f\mid J^{\mu}(\q) \mid \Psi_i\rangle \, 
\langle \Psi_i \mid J^{\nu\dagger}(\q)\mid \Psi_f\rangle \, 
\delta(E_i+\omega-E_f),
\label{eq.wmn}
\end{equation}
where $\omega$ and $\q$ are the energy and momentum transfer, respectively.
Different but consistent models to calculate the components of the hadron tensor 
in QE electron- and neutrino-nucleus scattering are outlined in the next 
sections.

\section{The exclusive $(e,e^{\prime}p)$ reaction}

For the exclusive $(e,e^{\prime}p)$ reaction nonrelativistic and relativistic 
models based on the distorted-wave impulse approximation (DWIA) have been 
developed to calculate the matrix elements in Eq.~(\ref{eq.wmn}). The DWIA 
expression of the matrix elements is the results of the following 
assumptions~\cite{rep,book,bof82}: 
\newline
i) An exclusive process is considered, where the residual nucleus is left in a
discrete eigenstate $n$ of its Hamiltonian.
\newline
ii) The final nuclear state is projected onto the channel subspace spanned by 
the vectors corresponding to a nucleon, at a given position, and the residual 
nucleus in the state $n$. This assumption neglects effects of coupled channels 
and is justified by the considered asymptotic configuration of the final state.
\newline
iii) The (one-body) nuclear-current operator does not connect different channel 
subspaces. Thus, also the initial state is projected onto the selected channel 
subspace. This assumption is the basis of the direct-knockout mechanism 
and is related to the IA.
  
The amplitudes of Eq.~(\ref{eq.wmn}) are then obtained in a one-body representation as 
\begin{equation}
\lambda_n^{1/2} \langle\chi^{(-)}\mid  j^{\mu}(\q)\mid \varphi_n \rangle  \ ,
\label{eq.dko}
\end{equation}  
where $j^{\mu}$  the one-body nuclear current, $\chi^{(-)}$ is the 
single-particle (s.p.) scattering state of the emitted nucleon, $\varphi_n$ the 
overlap between the ground state of the target and the final state $n$, 
i.e., a s.p. bound state, and the spectroscopic factor $\lambda_n$ is the norm
of the overlap function, that gives the probability of removing from the target
a nucleon leaving the residual nucleus in the state $n$. 
In the model the s.p. bound and scattering states are consistently derived as 
eigenfunctions of a non Hermitian energy dependent Feshbach-type optical 
potential and of its Hermitian conjugate at different energies. 
In standard DWIA calculations phenomenological ingredients are usually employed: 
the scattering states are eigenfunctions of a phenomenological optical potential 
determined through a fit to elastic nucleon-nucleus scattering data and the 
s.p. bound states are obtained from mean-field potentials, or can be calculated 
in a phenomenological Woods-Saxon well.  

The model can be formulated in a similar way within nonrelativistic 
DWIA and relativistic RDWIA frameworks~\cite{meucci01}. In RDWIA, calculations 
are performed with a relativistic nuclear-current operator and four-vector 
relativistic wave functions for the s.p. bound and scattering states. 
Both DWIA and RDWIA models have been quite successful in describing  a large 
amount of $(e,e^{\prime}p)$ data in a wide range of nuclei and in different
kinematics~\cite{book,meucci01,ud93,epja}. The shape of the experimental 
recoil-momentum distributions corresponding to a particular state of the 
residual nucleus are well described by model calculations. Then, the spectroscopic 
factors are usually extracted from the comparison between experimental and 
theoretical results as the reduction factors that must be applied to the 
calculated cross sections to reproduce the magnitude of the experimental cross 
sections.

Recently, the nonrelativistic DWIA and relativistic RDWIA models widely and
successfully adopted to describe data in a wide range of stable nuclei have been 
applied to nuclei with neutron excess, with the aim to study the evolution of 
the  $(e,e^{\prime}p)$ cross sections as a function of the asymmetry between 
the number of neutrons and protons~\cite{exot}. 

\section{Semi-inclusive neutrino-nucleus scattering}

For the semi-inclusive NC and CC processes  where only the outgoing nucleon is 
detected, the transition amplitudes can be calculated in the same RDWIA model
of Eq.~\ref{eq.dko}. Since the outgoing lepton is not detected, the cross section
must be integrated over the energy and angle of the outgoing lepton. The 
state of the residual nucleus is not determined and the cross section includes
all the states $n$.
In the calculations \cite{nc,nc1,spin-nc} a pure shell-model (SM) description is assumed, 
i.e., $n$  is a one-hole state and the sum over $n$ is over all the occupied 
SM states. In \cite{nc,nc1,spin-nc} FSI  are described by a complex optical potential whose 
imaginary part gives an absorption that reduces the calculated cross section. 
A similar reduction is found in $(e,e^{\prime}p)$ calculations. 
The imaginary part of the optical potential accounts for the fact that in the 
elastic scattering, if other channels are open besides the elastic one, part of
the incident flux is lost in the elastically scattered beam and goes to the 
inelastic channels that are open. In the exclusive scattering, where only
one channel is considered, it is correct to include the absorptive imaginary 
part of the optical potential and account for the flux lost in the considered 
channel, but in the inclusive scattering, where all the channels are included,
the flux lost in a channel must be found in the other channels, and in the sum 
over all the channels the flux can be redistributed but must conserved. 
Thus, the use of the absorptive imaginary part of the optical potential in the 
inclusive scattering seems inconsistent with the requirement of flux 
conservation. 
For the semi-inclusive process where the emitted nucleon is detected, some of 
the reaction channels which are responsible for the imaginary part of the 
potential like multistep processes, fragmentation of the nucleus, etc., are not 
included in the experimental cross section.  
There are,  however, contributions that are not considered in this model and 
that can be contained in the experimental cross section. 
In \cite{nc,nc1,spin-nc} the semi-inclusive neutrino scattering is treated as 
a process where the cross section is obtained from the sum of all the 
integrated exclusive one-nucleon knockout channels. We can expect that 
the description of FSI by means of a complex optical potential that gives 
absorption can produce cross sections that are smaller than the experimental 
data. However, since measurements of cross sections are a hard
experimental task, ratios of cross sections have been proposed as alternative 
quantities that can provide useful informations. In  \cite{nc,nc1,spin-nc} it 
is shown that ratios of cross sections are less sensitive to distortion effects 
and several ratios are considered in view of the possible determination of the 
strange quark content of the nucleon.   

\section{Inclusive lepton-nucleus scattering}

For the inclusive scattering where only the outgoing lepton is detected the 
Green's function (GF) model  \cite{eenr,ee,cc,eea,eesym,acta,acta1} has been developed   
to describe FSI consistently with the exclusive 
scattering and using the same complex optical potential.

In the GF model, under suitable approximations, that are basically related to the
IA, the components of the hadron tensor are written in terms of the s.p. 
optical model Green's function, whose self-energy the Feshbach optical 
potential. The explicit calculation of the s.p. Green's function can be 
avoided by its spectral representation, that is based on a biorthogonal 
expansion in terms of a non Hermitian optical potential $\cal H$ and of its 
Hermitian conjugate $\cal H^{\dagger}$. Calculations require matrix elements 
of the same type as the DWIA ones in Eq. \ref{eq.dko}, but involve 
eigenfunctions of both $\cal H$ and $\cal H^{\dagger}$, where the imaginary 
part gives in one case an absorption and in the other case a gain of flux, and 
in the sum over $n$ the total flux is redistributed and conserved.  
The GF approach guarantees a consistent treatment of FSI in the exclusive and 
in the inclusive scattering and gives a good description of $(e,e')$ 
data~\cite{eenr,ee}.

\begin{figure}
\begin{center}
\includegraphics[width=90mm]{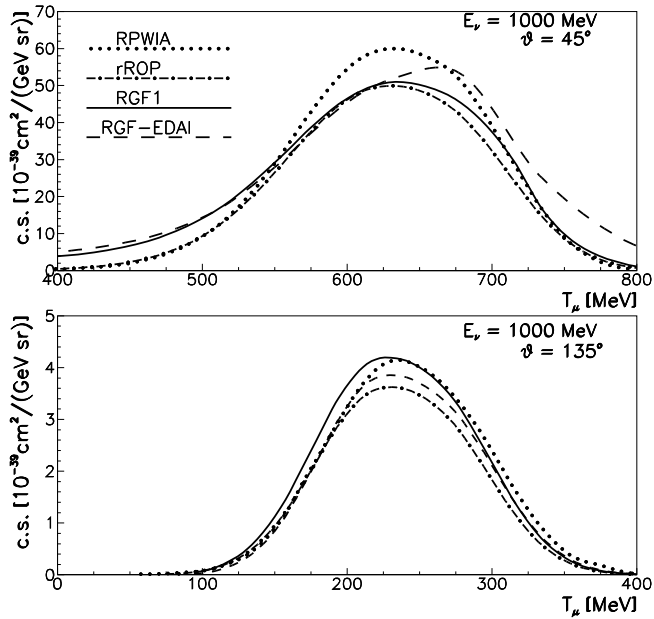}
\end{center}
\vskip  -3mm
\caption{The cross sections of the $^{12}$C$(\nu_{\mu},\mu^-)$ 
reaction for an incident neutrino energy of $E_\nu$ = 1000 MeV and a muon
scattering angle $\vartheta = 45^{\mathrm o}$ (upper panel) 
and $135^{\mathrm o}$ (lower panel) as a
function of the muon kinetic energy $T_\mu$.  
Results for RPWIA (dotted), rROP (dot-dashed), and the relativistic RGF 
approach with  two different optical potentials \cite{chc}, i.e., EDAD1 (RGF1) 
and EDAI (RGF-EDAI), are compared. 
\label{fig1}}
\end{figure}

For the inclusive electron scattering both nonrelativistic GF~\cite{eenr,eesym} 
and relativistic RGF~\cite{ee} approaches have been considered. For CC 
neutrino scattering the RGF model has been adopted. A numerical example is 
shown in Fig. \ref{fig1} for the $^{12}$C$(\nu_{\mu},\mu^-)$  cross 
sections calculated for an incident neutrino energy of $E_\nu$ = 1000 MeV and 
two values of the muon scattering angle, {\it i.e.}, 
$\vartheta = 45^{\mathrm o}$ (upper panel) and $135^{\mathrm o}$ (lower panel), 
as a function of the muon kinetic energy $T_\mu$. In the case of the RGF model
two different parameterizations of the relativistic optical potential have been
used in the calculations, {\it i.e.}, the energy-dependent and A-dependent 
EDAD1 (RGF1) and the energy-dependent but A-independent EDAI (RGF-EDAI), which are fitted to proton elastic scattering data on several nuclei in an energy range up to 
1040 MeV \cite{chc}. In Fig. \ref{fig1} the RGF1 and RGF-EDAI results are compared 
with the results of the relativistic 
plane wave IA (RPWIA), where FSI are neglected, and with the cross sections 
obtained when only the real part of the relativistic optical potential (rROP) 
is retained and the imaginary part is neglected.  
The rROP approximation conserves the flux but, independently of its numerical
results, it is conceptually wrong because the optical potential has to be complex 
owing to the presence of inelastic channels.
 
The differences between the rROP, RGF1, and RGF-EDAI results are essentially due to 
the imaginary part of the optical potential. Different parameterizations of the
optical potential are available, that are able to give equivalently good
descriptions of elastic proton-nucleus scattering data, but correspond to
different optical potentials that mainly differ for their imaginary part.
The imaginary part accounts for the overall effects of inelastic channels and
is not univocally determined from the elastic phenomenology. 
The real terms are very similar and the cross sections 
calculated in rROP are practically insensitive to the choice of the optical 
potential. In contrast, the imaginary part is sensitive to the 
parameterization of the ROP and gives the differences shown in the figure. 
The RGF1 and RGF-EDAI results at $\vartheta = 45^{\mathrm o}$ have somewhat 
different shapes for high values of $T_\mu$, {\it i.e.}, low values of the 
energy transferred. At $\vartheta = 135^{\mathrm o}$ the differences are 
reduced but the magnitude of the cross sections is significantly reduced.

\section{Comparison of relativistic models}

The analysis of data for neutrino experiments requires a precise knowledge of 
lepton-nucleus cross sections, where uncertainties on nuclear effects are 
reduced as much as possible. To this aim, it is important to check the 
consistency of different models and the validity of the adopted approximations.   
The results of the relativistic models developed by the Pavia and the
Madrid-Sevilla groups for the inclusive scattering have been compared
in~\cite{compee} for the inclusive electron scattering and in~\cite{compcc} for
the inclusive CC neutrino scattering. 
 As a first step, the consistency of the RPWIA and rROP calculations performed 
by the two groups with independent  numerical programs and with the same
theoretical ingredients has been checked. 
Almost identical results are obtained in RPWIA and very similar results in rROP.
This result gives us confidence on the reliability of our calculations and 
allows us to extend the comparison to the different descriptions of FSI. 

\begin{figure}
\begin{center}
\includegraphics[width=90mm]{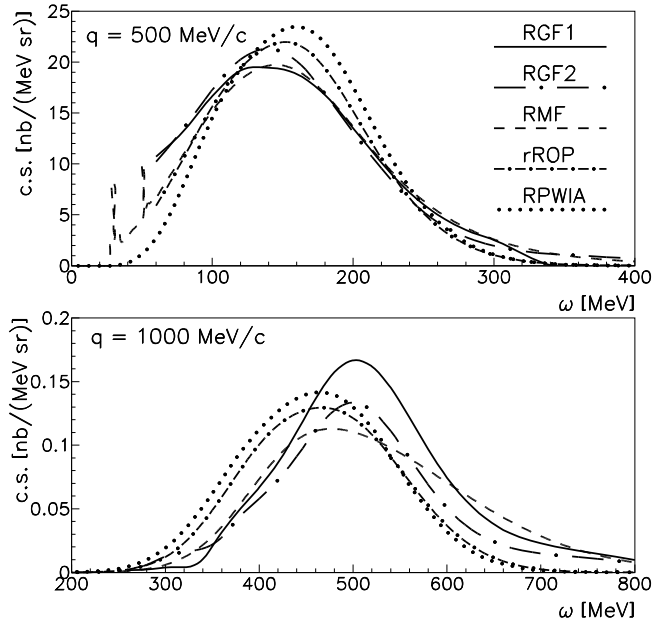}
\end{center}
\vskip  -3mm
\caption{Differential cross section of the 
$^{12}$C$(e,e^{\prime})$ reaction for an incident electron energy 
$\varepsilon = 1$ GeV and $q = $ 500 and 1000 MeV$/c$. Results for RPWIA (dotted), rROP (dot-dashed), RGF1 (solid),  
RGF2 (long dot-dashed), and RMF (dashed) are compared.
\label{fig2}}
\end{figure}
An example is shown in  Fig. \ref{fig2}, where the  $^{12}$C$(e,e')$ cross 
sections obtained by the two groups in RPWIA and rROP are compared with the 
RGF results obtained with two different energy-dependent and 
A-dependent parameterizations for the 
relativistic optical potential of~\cite{chc}, {\it i.e.}, the EDAD1 (RGF1), already considered for the calculations shown in  
Fig.~\ref{fig1}, and the 
EDAD2 (RGF2), 
and with the results of the relativistic mean  field (RMF) model \cite{cab} of 
the Madrid-Sevilla group, where the scattering wave functions are calculated 
with the same strong real mean-field potential used for the initial bound 
states. The RMF model fulfills the dispersion relation and maintains 
the continuity equation.

The differences between the RMF and RGF results increase with the momentum 
transfer. Also the discrepancies between the RGF1 and RGF2 cross sections depend 
on the momentum transfer.  At $q$ = 500 MeV$/c$ the three results are similar, 
both in magnitude and shape, larger differences are obtained 
at $q = 1000$ MeV$/c$. The shape of the RMF cross section shows an asymmetry, 
with a long tail extending towards higher values of $\omega$, that is essentially due to the 
strong energy-independent scalar and vector potentials present in the RMF 
model.
The asymmetry towards higher $\omega$ is less significant but still visible 
for RGF1 and RGF2, whose cross sections show a
similar shape but with a significant difference in the magnitude. 
At $q$ = 1000 MeV$/c$ both RGF1 and RGF2 cross sections are higher than the RMF 
one in the region where the maximum occurs.
A stronger enhancement is obtained with RGF1, which at the peak 
overshoots the RMF cross section up to $40\%$ and it is even higher than the
RPWIA result.

The behavior of the RMF and RGF results as a function of $q$ and
$\omega$  is linked to the structure of the relativistic 
potentials involved in the RMF and RGF models. Whereas RMF is based on the 
use of a strong energy-independent real potential, RGF makes use 
of a complex energy-dependent optical potential. In RGF calculations the 
behavior of the optical potential changes with the momentum and energy 
transferred in the process, and higher values 
of $q$ and $\omega$ correspond to higher energies for the optical potential. 
The RGF results are consistent with the general behavior of the 
optical potentials and are basically due to their imaginary part. 
The real terms of the relativistic optical potentials are very similar for the 
different parameterizations and the rROP cross sections do not show 
sensitivity to the particular parameterization considered. On the other hand, the 
energy-dependent scalar and vector components of the real part of the ROP get 
smaller with increasing energies. Thus the rROP result approaches the RPWIA one 
for large values of $\omega$. In contrast, the imaginary (scalar and vector) 
part presents its maximum strength around 500 MeV, being also sensitive to the 
particular ROP parameterization. This explains the differences observed
between the rROP and the two RGF results as a function of $\omega$ and $q$.

\begin{figure}
\begin{center}
\includegraphics[width=100mm]{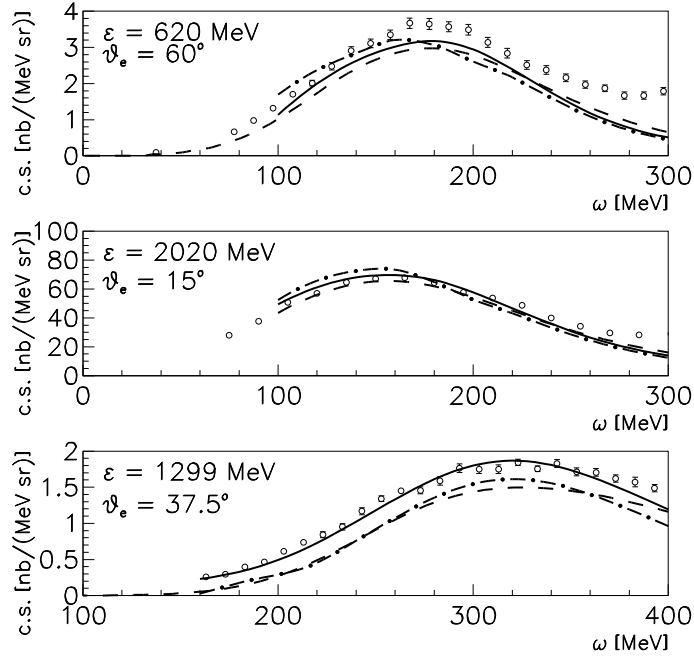}
\end{center}
\vskip  -3mm
\caption{Differential cross section of the 
$^{12}$C$(e,e^{\prime})$ reaction for different beam energies and 
electron-scattering angles in RGF1 (solid), RGF2 (long dot-dashed), RMF (dashed). 
Experimental data from~\cite{620,day,1299}. 
} 
\label{fig3}
\end{figure}
In Fig.~\ref{fig3} the RGF1, RGF2, and RMF results are compared with the
experimental cross sections for three different kinematics~\cite{620,day,1299}. 
The three models lead to similar cross sections. The main differences are 
obtained for higher values of the momentum transfer, about 800 MeV/$c$ (bottom panel),
where the RGF1 cross section (solid line) is larger than the RGF2 (dot-dashed) 
and RMF (dashed) ones. The experimental cross section is slightly underpredicted 
in the top panel and well described in the middle panel by all calculations. 
Finally, results in the bottom panel show a fair agreement with data in the case of
RGF1, whereas RGF2 and RMF underpredict the experiment. Although satisfactory on 
general grounds, the comparison with data gives only an indication and cannot 
be conclusive until contributions beyond the QE peak, like meson-exchange 
currents and Delta effects, which may play a significant role in the analysis of data 
even at the maximum of the QE peak, are carefully 
evaluated~\cite{BCDM04,amaro05,ivanov08}. 

\begin{figure}
\begin{center}
\includegraphics[width=90mm]{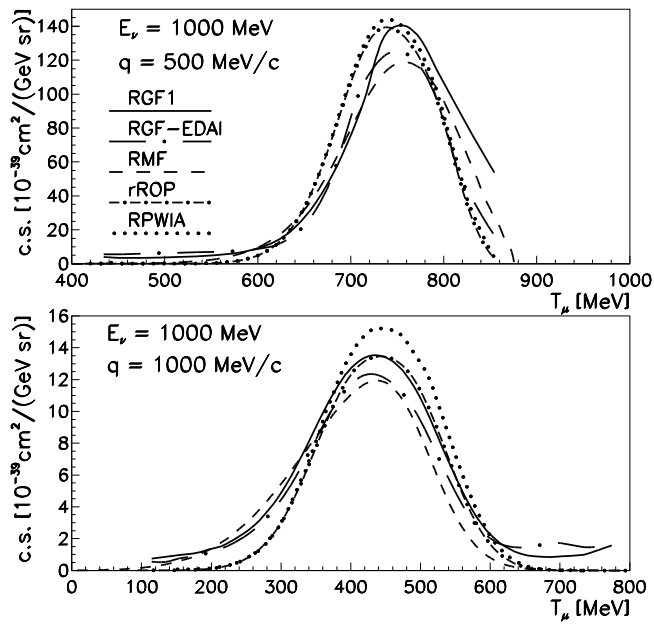}
\end{center}
\vskip  -3mm
\caption{Differential cross section of the 
$^{12}$C$(\nu_{\mu} , \mu ^-)$ reaction 
	for E$_{\nu}$ = 1000 MeV and $q$ = 500 MeV/$c$ and 1000 MeV/$c$.
 Results for RPWIA (dotted), rROP (dot-dashed) RGF1 (solid), RGF-
 EDAI (long dot-dashed), and RMF (dashed) are compared.
\label{fig4}}
\end{figure}
In Fig.~\ref{fig4} the $^{12}$C $(\nu_{\mu},\mu^{-})$ cross section
obtained by the two groups in RPWIA and rROP in a kinematics with a fixed value 
of the incident neutrino energy, $E_{\nu} = 1000$ MeV,  and two values of the 
momentum transfer, {\it i.e.}, $q$ = 500 and 1000 MeV$/c$, are compared with the
corresponding RMF, RGF1 and RGF-EDAI results. 

The cross sections shown in Fig.~\ref{fig4} are calculated with same incident 
lepton energy and momentum transfer as in the $(e,e^{\prime})$ 
calculations of Fig.~\ref{fig2}. This is a kinematics more typical of 
electron-scattering experiments than of neutrino experiments, but the 
calculations have been carried out to perform a more direct comparison between the results of the
models in electron and neutrino scattering. Actually, in the case of CC 
neutrino scattering the muon mass gives a different kinematics, with different 
values of the energy transfer and, as a consequence, of the energies of the 
outgoing nucleon. 
This means that in the RGF model the optical potential is calculated for 
electron and neutrino scattering at different energies. We have checked that 
if we reproduce the kinematics of electron scattering and perform 
calculations for the $(\nu_\mu , \mu ^-)$ reaction with vanishing muon mass, the 
main difference with respect to the calculations shown in Fig.~\ref{fig4} is 
a shift of the cross section by about 100 MeV towards higher values of 
$T_{\mu}$, without any significant change in the shape or in the strength. 

Also for the results in Fig.~\ref{fig4} the shape of the RMF cross section 
shows an asymmetry with a tail extending towards higher values of $\omega$
(corresponding to lower values of the kinetic energy of the outgoing muon
$T_\mu$). An asymmetric shape towards higher $\omega$ is shown also by the RGF 
cross sections, while no visible asymmetry is given by the RPWIA and rROP 
results. Also in this case the significant differences obtained between the RGF 
and rROP cross sections are consistent with the general 
behavior of the phenomenological energy-dependent relativistic optical 
potentials and are basically due to their different imaginary 
part.
As already shown for $(e,e')$ reactions, the RGF yields in general a larger 
cross section than the RMF. The RGF and the RMF yield similar predictions, 
within a few percent for low-$q$, while as $q$ goes up the RGF yields 
increasingly larger cross sections than RMF. This may reflect the influence of 
the pionic degrees of freedom, that can be included in a phenomenological way in
the imaginary part of the optical potential \cite{compee,compcc}.

The results in Fig.~\ref{fig4} present some differences with respect to the 
corresponding cross sections of the inclusive electron scattering shown in 
Fig.~\ref{fig2}. In both cases the differences
between the results of the different models are generally larger for increasing value of 
the momentum transfer. For neutrino scattering, however, such a behavior is 
less evident and clear.  In particular, the RGF1 cross section at 
$q$ = 1000 Mev$/c$ does not show the strong enhancement in the region of the 
maximum that is found for the $(e,e^{\prime})$ calculations of 
Fig.~\ref{fig2}, where the RGF1 result is even larger than the RPWIA one. 
In the case of neutrino scattering the RGF results in the region of the 
maximum are generally larger than the RMF ones, but smaller than the RPWIA 
cross sections. 

In spite of many similarities, inclusive electron 
scattering and CC neutrino-nucleus scattering are two different processes and 
caution should be drawn on their comparison.
The different currents and their possible interplay with the other 
ingredients of the models do not allow an easy comparison.
The numerical differences between the RGF results for electron and neutrino 
scattering can mainly be ascribed to the combined effects of the weak current, 
in particular its axial term, and the imaginary part of the ROP. We have 
checked that these effects can explain the fact that in neutrino scattering the 
RGF1 result does not give the strong enhancement in the region of the maximum 
that is obtained for the $(e,e^{\prime})$ cross section in Fig.~\ref{fig2}.

\begin{figure}
\begin{center}
\includegraphics[width=90mm]{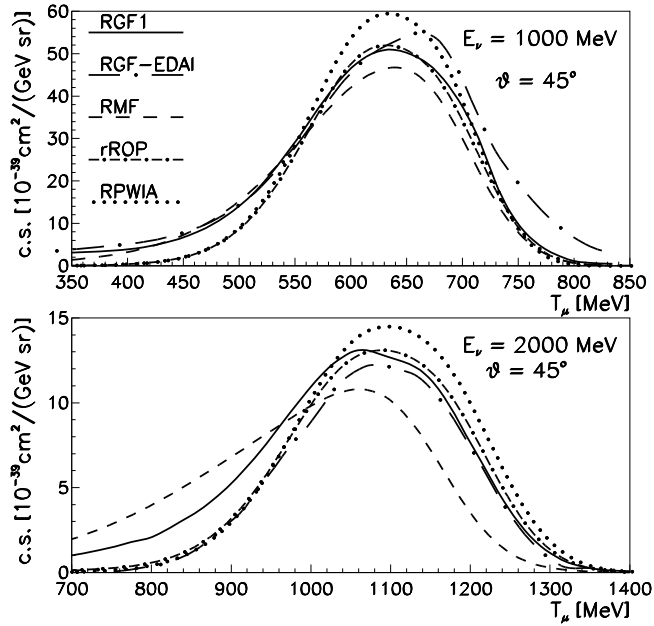}
\end{center}
\vskip  -3mm
\caption{Differential cross section of the 
$^{12}$C$(\nu_{\mu} , \mu ^-)$ reaction 
	for E$_{\nu}$ = 1000 MeV and 2000 MeV, and scattering 
	angle $\vartheta = 45^{\mathrm o}$.
 Results for RPWIA (dotted), rROP (dot-dashed) RGF1 (solid), RGF-EDAI (long dot-dashed), and RMF (dashed) are compared.
\label{fig5}}
\end{figure}
The $^{12}$C $(\nu_{\mu},\mu^{-})$ cross section for the  kinematics with  
$E_{\nu}$ = 1000 and 2000 MeV and $\vartheta = 45^{\mathrm o}$ are presented 
in Fig.~\ref{fig5}.
In these kinematics, that are more similar to those actually explored at 
present neutrino experiment facilities, the momentum transfer is not fixed and 
its value around the peak of the cross section is usually large, {\it i.e.}, 
$q \approx 700$ MeV$/c$ for $E_{\nu} = 1000$ MeV and $q \approx 1400$ MeV$/c$ 
for $E_{\nu} = 2000$ MeV.
The shape of the RMF cross section shows also in this case an asymmetry, with 
a long tail extending towards lower values of $T_\mu$, which is due to the 
strong energy-independent scalar and vector potentials present in the RMF 
approach. The asymmetry increases with larger incident neutrino energies. 
For the RGF cross sections the asymmetry is less significant but still visible. 
while almost no asymmetry is found for the RPWIA and rROP cross sections. 
The RGF1 and RGF-EDAI cross sections have somewhat different shapes, that are 
particularly visible for low $\omega$ at  $E_{\nu} = 1000$ MeV and for higher 
$\omega$ at $E_{\nu} = 2000$ MeV. These differences are essentially due
to the imaginary part of the ROP, that is sensitive to the particular 
parametrization considered, while the real terms of the ROP's are 
very similar for different parameterizations and give very similar results. 

\section{Comparison with Charged-Current Quasielastic MiniBooNE data}

The double-differential cross sections for muon neutrino CC 
quasielastic (CCQE) scattering have  recently been measured by the MiniBooNE 
collaboration~\cite{miniboone}. These data have raised a strong debate over 
the role of the theoretical ingredients entering the description of the 
reaction. The experimental cross section is underestimated by the relativistic 
Fermi Gas (RFG) model, and also by the results of other more sophisticated
models,  unless the value of nucleon axial mass $M_A$ is significantly 
enlarged (1.35 GeV/$c^2$ in the RFG) with respect to the accepted world average 
value (1.03 GeV/$c^2$~\cite{Bern02,bodek}).
Before drawing conclusions about the need to increase the axial mass it is 
however important to evaluate carefully the contributions of all nuclear
effects. 

Within the QE kinematic domain the treatment of FSI is essential for
the comparison with data. The comparison between the results of the RMF and RGF 
models, where FSI are described by very different ingredients, can be helpful 
for a deeper understanding of the role played by FSI in the 
analysis of CCQE data and its influence in studies of neutrino oscillations at 
intermediate to high energies. The predictions of the two models have been 
compared with the recent CCQE MiniBooNE data \cite{compmini}.

\begin{figure}
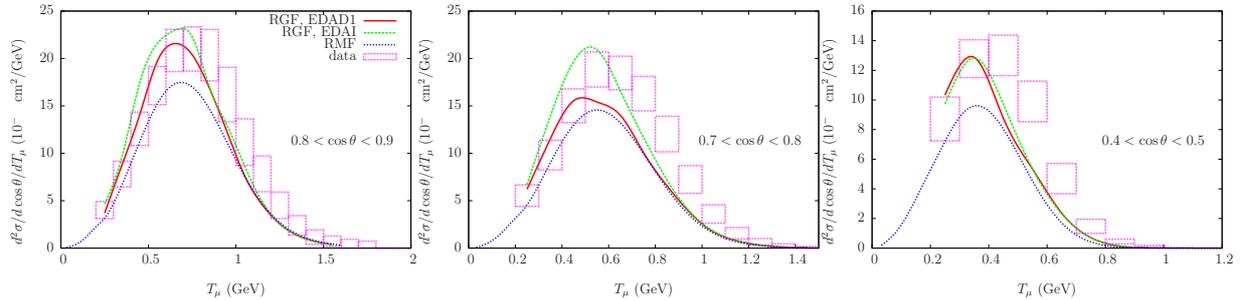

\begin{center}
\includegraphics[scale=0.45]{COS085.epsi}%
\includegraphics[scale=0.45]{COS075.epsi}%
\includegraphics[scale=0.45]{COS045.epsi}%
\end{center}
\vskip  -3mm
\caption{Flux-averaged double-differential cross section per 
target nucleon for the CCQE $^{12}$C$(\nu_{\mu} , \mu ^-)$ reaction calculated 
in the RMF (blue line), the RGF1 (red), and 
	RGF-EDAI (green) and displayed versus $T_\mu$ for various bins 
	of $\cos\theta$. In all the calculations the standard value of the 
	nucleon axial 
mass, {\it i.e.}, $M_A = 1.03$ GeV/$c^2$ has been used.
The data are from MiniBooNE~\cite{miniboone}.
\label{fig6}}
\end{figure}
The CCQE double-differential $^{12}$C $(\nu_{\mu},\mu^{-})$ cross sections 
averaged over the neutrino flux as a function of $T_\mu$ for various bins 
of $\cos\theta$, where $\theta$ is the muon scattering angle, are shown in 
Fig.~\ref{fig6}.
The RMF results yield reasonable agreement with data for small angles and low muon energies, 
the discrepancy becoming larger as $\theta$ and $T_\mu$ increase. The shape 
followed by the RMF cross sections fits well the slope shown by the data. 
A good agreement with the experimental shape is shown also by 
the RGF cross sections. The RMF and RGF models yield close predictions at
larger values of $T_\mu$, for all the bins of $\cos\theta$ shown in 
the figure. The RGF cross sections are generally larger than the RMF ones. 
The differences increase approaching the peak region, where the additional
strength shown by the RGF produces cross sections in reasonable agreement 
with the data. 

Also the differences between the RGF results with the two optical potentials  
are enhanced in the peak region, but remain in general within the 
uncertainties of the experimental errors. The EDAD1 and EDAI potentials yield 
close predictions for the bin $0.4<\cos\theta<0.5$, the differences are visible
but anyhow small for the bin $0.8<\cos\theta<0.9$, being the RGF-EDAI cross 
section larger than the RGF1 one, while the difference is sizeable for the bin 
$0.7<\cos\theta<0.8$, with the RGF1 result closer to the RMF than to the 
RGF-EDAI one.

\begin{figure}
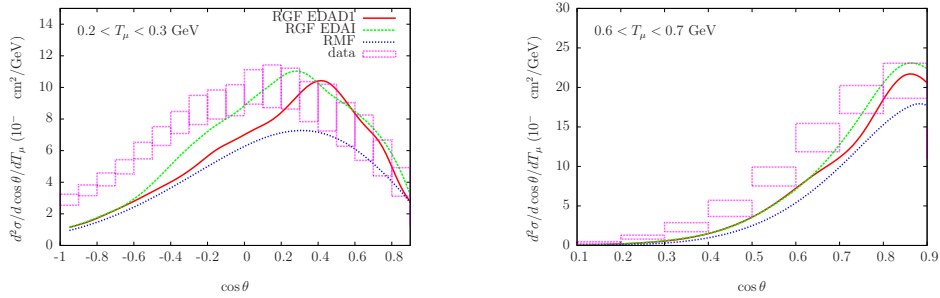

\begin{center}
\includegraphics[scale=0.45]{T025_smooth.epsi}%
\hskip  1.5cm
\includegraphics[scale=0.45]{T065_smooth.epsi}%
\end{center}
\vskip  -3mm
\caption{Flux-averaged double-differential cross section per 
target nucleon for the CCQE $^{12}$C$(\nu_{\mu} , \mu ^-)$ reaction displayed
versus $\cos\theta$ for two bins of $T_\mu$.  The results obtained with 
RMF (blue line), RGF1 (red), and RGF-EDAI (green) are compared. 
In all the calculations the standard value of the 
	nucleon axial 
mass, {\it i.e.}, $M_A = 1.03$ GeV/$c^2$ has been used.
The data are from MiniBooNE~\cite{miniboone}.
\label{fig7}}
\end{figure}
In Fig.~\ref{fig7} the flux-averaged double-differential cross sections
are plotted versus $\cos\theta$ for two bins of $T_\mu$, {\it i.e.}, 
$0.2<T_\mu<0.3$ GeV and  $0.6<T_\mu<0.7$ GeV. The shape of the experimental
cross section is well described by the models. The RMF results generally
understimate the data, especially for the lower muon energy values, the 
agreement improves as $T_\mu$ increases. The additional strength produced by the complex optical potential in the RGF 
gives a reasonable agreement with the size of the experimental cross 
section. The agreement is better for smaller angles while the data are slightly 
underpredicted as $\theta$ increases.
The discrepancy is larger with RGF1, that gives in general a lower cross 
section than RGF-EDAI, and in the bin $0.2<T_\mu<0.3$ GeV.

These results give further and clear indication that before 
drawing conclusions about the comparison with data and about the need to 
increase the value of the nucleon axial mass a careful evaluation of all nuclear 
effects is required. Important contributions can be produced by FSI. Both RGF 
and RMF models, where FSI are described with very different theoretical
ingredients, give a good description of the shape of the experimental cross 
sections. The RMF generally underpredicts the data, but for lower values of the
muon energy and scattering angle. In contrast, the RGF model can give cross 
sections of the same magnitude as the experimental ones without the need to 
increase the standard value of the axial mass. 
The larger cross sections in the RGF model are produced by the imaginary part 
of the ROP, that includes the overall effect of inelastic (nucleonic and
non-nucleonic) channels and that is not univocally determined from the elastic 
phenomenology.
The choice of the optical potential and a more refined determination of its 
imaginary part can deserve further investigation. 
Other contributions, not included in the present calculations, might play an
important role  in the analysis of CCQE data. Results of different models
indicate that significant effects can be expected from correlations and
meson-exchange currents \cite{amaro11a,amaro11b} or multiple knockout
excitations \cite{Martini,Martini1,Nieves11,Nieves11b}. 
More refined calculations including all these contributions should be 
performed, but any new contribution should be included consistently in a model. 
 
\ack

The authors wish to acknowledge the collaborations that have produced the 
results reported in this contribution. In particular, we are grateful to 
Franco Pacati, Franco Capuzzi, Juan Antonio Caballero, Jos\'{e} Manuel  
Ud\'{\i}as, and Maria Barbaro for fruitful collaborations.

\end{document}